\documentclass[11pt]{aastex}
\usepackage{graphics}
\usepackage{natbib}

\title{UV Photolysis, Organic Molecules in Young Disks, and the Origin of Meteoritic Amino Acids \\ \ \\ 
Submitted to \textsl{Icarus} 14-Jun-2010\\
Revised 13-Dec-2010\\
Accepted 4-Jan-2011\\}
\shorttitle{UV and Organics in Disks}
\shortauthors{Throop}

\begin{document}



%


\def \ApJ	 	{Astrophys. J.}
\def \apj 		{Astrophys. J.}
\def \apjl 		{Astrophys. J. Lett.}
\def \apjs 		{Astrophys. J. Suppl.}
\def \AJ 		{Astron. J.}
\def \aj 		{Astron. J.}
\def \aap 		{Astron. \& Astrophys.}
\def \aaps 		{Astron. \& Astrophys. Suppl.}
\def \apss 		{Astrophys. \& Spac. Sci.}
\def \mnras 		{Month. Not. Royal Astron. Soc.}
\def \nat 		{Nature}
\def \grl 		{Geophys. Res. Let.}
\def \jgr 		{J. Geophys. Res.}
\def \mps 		{Meteorit. Plan. Sci.}
\def \pasp 		{Pub. Astron. Soc. Pac.}
\def \planss 		{Plan. Spac. Sci.}
\def \baas 		{Bull. Amer. Astron. Soc.}
\def \ao		{Appl. Optics}
\def \araa		{Ann. Rev. Astron. Astrophys.}
\def \jqsrt		{J. Quant. Spect. Rad. Trans.}
\def \ssr		{Space Sci. Rev.}


\def \cms		{ cm s$^{-1}$ }			
\providecommand{\kms}   {\ensuremath{\rm{km\ s^{-1}}}}
\def \gcm		{ g cm$^{-2}$ }			
\def \per		{ $^{-1}$ }			
\def \pertwo		{ $^{-2}$ }			
\providecommand{\sol}	{\ensuremath{{\odot}}}		
\providecommand{\msol}	{\ensuremath{M_{\sol}}}		
\providecommand{\lsol}	{\ensuremath{L_{\sol}}}		
\providecommand{\esol}	{\ensuremath{E_{\sol}}}		
\providecommand{\mjup}	{\ensuremath{M_{\rm{J}}}}	
\providecommand{\mj}	{\ensuremath{M_{\rm{J}}}}	
\def \water		{H$_2$O}			
\providecommand{\tonec}		{\ensuremath {$\theta^1$ Ori C} }		
\newcommand \isotope[2] {\ensuremath{{}^{#2}\rm{#1}}}   

\def \paa		{Pa$\alpha$}			
\def \ha		{H$\alpha$}
\def \Ha		{H$\alpha$}
\def \hii		{H{\small II}}
\def \sii		{S{\small II}}
\def \oiii		{O{\small III}}
\def \arcmin		{'}
\def \eqtext#1		{\hspace{1in} \hbox{#1}}	
\providecommand{\micron}{\ensuremath{\mu\rm{m}}}
\providecommand{\tpp}   {\ensuremath{\tau \varpi_0 P}}	
\providecommand{\rgi}	{\ensuremath{\rm{r}_{gI}}}
\providecommand{\rgii}	{\ensuremath{\rm{r}_{gII}}}
\def \etal      	\hbox{ \it et al.} 		
\def \etals		\hbox{{\it et al.}'s}		
\def \vs		\hbox{vs.}			
\providecommand{\degrees}{\ensuremath{{}^\circ}}
\providecommand{\mearth} {\ensuremath{M_{\oplus}}}	
\def \b			{$\bullet\ $}
\def \Beta		{\beta}				
\def \yes        	{$\surd\ $}     		
\def \idlplot#1		{\centerline{\scalebox{0.9}{\includegraphics{#1}}}} 
\def \idlplotps#1	{\centerline{\scalebox{0.9}{\includegraphics[70,350][574,710]{#1}}}} 
\def \ie		{{\it i.e.\/}}
\def \eg		{{\it e.g.\/}}
\def \subsimt#1{{\lower 2pt\hbox{$\scriptstyle #1$}\atop
     \raise 1pt\hbox{$\scriptstyle \sim$}}}
\def \gtrsim    	{\subsimt >}			
\def \lessim    	{\subsimt <}
\def \lesssim    	{\subsimt <}

\providecommand{\gt} 	{\ensuremath{>}}
\providecommand{\lt} 	{\ensuremath{<}}


\newcommand{\rj}	{\ensuremath{R_{\hbox{J}}}}


\newcommand{\rbh}	{\ensuremath{R_{\rm{B}}}}
\newcommand{\rtidal}	{\ensuremath{R_{\rm{T}}}}
\newcommand{\racc}	{\ensuremath{R_{\rm{Acc}}}}
\newcommand{\dmdt}	{\ensuremath{\dot{M}}}
\newcommand{\dmdtbh}	{\ensuremath{\dot{M}_{\rm{B}}}}
\newcommand{\mdotbh}	{\dmdtbh}
\newcommand{\dmbh}	{\ensuremath{{\Delta M}_{\rm{B}}}}
\newcommand{\dmdtacc}	{\ensuremath{\dot{M}_{\rm{acc}}}}
\newcommand{\dmdtstar}	{\ensuremath{\dot{M}_{*}}}
\newcommand{\msolyr}	{\ensuremath{M_{\odot}\ \rm{yr^{-1}}}}
\newcommand{\msolmyr}	{\ensuremath{M_{\odot}\ \rm{Myr^{-1}}}}
\newcommand{\mmsnmyr}	{\ensuremath{\rm{MMSN}\ \rm{Myr^{-1}}}}
\newcommand{\mdot}	{\ensuremath{\dot{M}}}


\newcommand{\promille}{ 
  \relax\ifmmode\promillezeichen
        \else\leavevmode\(\mathsurround=0pt\promillezeichen\)\fi}
\newcommand{\promillezeichen}{%
  \kern-.05em%
  \raise.5ex\hbox{\the\scriptfont0 0}%
  \kern-.15em/\kern-.15em%
  \lower.25ex\hbox{\the\scriptfont0 00}}


\def \sk		{\vskip 0.1 in}

\def \doublespace 	{\baselineskip = 24 pt}
\def \singlespace 	{\baselineskip = 12 pt}
\def \halfspace 	{\baselineskip = 18 pt}

\font\bigrm = 		cmr10 scaled \magstep 1
\font\bigbigrm = 	cmr10 scaled \magstep 2
\font\halfbigrm = 	cmr10 scaled \magstephalf
\font\bigbf = 		cmb10 scaled \magstep 1
\font\bigbigbf = 	cmb10 scaled \magstep 2
\font\halfbigbf = 	cmb10 scaled \magstephalf

\def \in		{\leftskip = 0.0 in}
\def \ini	 	{\leftskip = 0.2 in}		
\def \inii	 	{\leftskip = 0.4 in}
\def \iniii	 	{\leftskip = 0.6 in}
\def \iniiii	 	{\leftskip = 0.8 in}

\def \hang 		{\parindent = -0.15 in \leftskip = 0.15 in}	
\def \nohang		{\parindent = 0 in \leftskip = 0 in}

\def \figstart		{\leftskip = 0.5 in \rightskip = 0.5 in}	
\def \figend		{\leftskip = 0 in \rightskip = 0 in}	

\def \instart		{\leftskip = 0.2 in \rightskip = 0.2 in}	
\def \inend		{\leftskip = 0 in \rightskip = 0 in}



\author{Henry B. Throop}
\affil{Southwest Research Institute}
\affil{Department of Space Studies}
\affil{1050 Walnut St Ste 300, Boulder, CO  80302}
\email{throop@boulder.swri.edu}

\begin{abstract}

The origin of complex organic molecules such as amino acids and their precursors found in meteorites and comets is
unknown.  Previous studies have accounted for the complex organic inventory of the Solar System by aqueous chemistry on
warm meteoritic parent bodies, or by accretion of organics formed in the interstellar medium.  This paper proposes
a third possibility: that complex organics were created \textit{in situ} by ultraviolet light from nearby O/B stars
irradiating ices already in the Sun's protoplanetary disk.  If the Sun was born in a dense cluster near UV-bright stars,
the flux hitting the disk from external stars could be  many orders of magnitude higher than that from the Sun alone.
Such photolysis of ices in the laboratory can rapidly produce amino acid precursors and other complex organic molecules.
I present a simple model coupling grain growth and UV exposure in a young circumstellar disk.  It is shown that the
production may be sufficient to create the Solar System's entire complex organic inventory within $10^6\ \rm{yr}$.
Subsequent aqueous alteration on meteoritic parent bodies is not ruled out.

\end{abstract}

\section{Introduction}

The early Solar System was replete with complex organic molecules, which can be seen today in preserved ancient bodies
such as meteorites and comets.  Upwards of 100 different amino acids have been detected on chondritic meteorites such as
Murchison, Allende and Orguiel.  Many of these have no known natural terrestrial occurrence \citep{egb01}, and are
believed to be of extra-terrestrial origin.  Comets and the interstellar medium (ISM) are also rich
inventories of complex organic molecules, including amino acid precursors \citep{sny06}.  Amino acids and other complex
organics have have implications for the origins of life on Earth, so understanding their formation and history is an
area of great interest.  If these molecules are formed and distributed easily in a variety of disks and conditions, then
pre-biotic compounds may be common throughout distant planetary systems.

Of the complex organics in the solar system, amino acids are particularly interesting to study.  These compounds are
necessary for life, and may have been biotic precursors on Earth.  Carbonaceous chondrites are typically 5\% or more
carbon by mass, most of which is in aromatic polymers and thousands of other pre-biotic organic compounds
\citep{sgg10,ha81}.  Laboratory studies have found Murchison to have some 10--30 ppm by mass in identified amino acids
\citep{ss90b}.  The simplest amino acid (glycine, NH$_2$CH$_2$COOH) has recently been detected in comets, but so far has
eluded detection in the ISM \citep{egd09,sny06,kch03}.  Potential amino precursors such as formic acid (HCOOH) have been
detected in comets but not the interstellar medium \citep{hps03,blw00}.  However, amino acids in ice form in the
interstellar medium (ISM) would be more stable but harder to detect than their corresponding gas forms in comets
\citep{ebd01,ctb90}, so it is likely that they exist in the ISM but have not yet been detected.

The origin of these molecules is unknown.  Two broad explanations exist for the amino acids in our solar
system.  First, they may be produced \textit{endogenically}, by chemical synthesis within the solar system itself.  
A variety of energy sources for this exist, including infall heating, radiogenic heating, lightning, and shocks.  For
the amino acids present on meteoritic parent bodies, the best-studied endogenic process is Strecker synthesis.  This is
a method by which amino acids are formed in aqueous environments, such as the warm sub-surface aquifers that could have
been present on asteroids as they were heated by $^{26}$Al and other radionuclides during their first few Myr
\citep{egb01}.  Amino acids are produced in the laboratory by this process, and sufficient $^{26}$Al existed to heat
parent bodies to liquid water temperatures \citep{mgg02}.  However, a serious problem exists with this model.  Isotopic
measurements of the chondritic amino acids consistently show deuterium \textit{enhancements} of $\delta D$ =
600-2000\promille, while measurements of the water in these same bodies show deuterium \textit{depletions} of roughly
100\promille.  \textit{In situ} synthesis of aminos does not preferentially change the D/H ratio, so this argues that
the water and organics come from distinct sources \citep{ler97}.

Second, the molecules may have been inherited \textit{exogenically} from the interstellar medium which formed
the solar nebula.  The ISM is known to be rich in complex chemistry, with over 150 different gas-phase species detected
to date in molecular clouds \citep{hv09}.   Amino acids have not been detected in the ISM, but sugars, alcohols,
polycyclic aromatic hydrocarbons (PAHs), and other complex molecules have been.  The formation of these compounds in the
ISM is thought to be due to a combination of processes, including gas phase, gas-grain, and UV-grain reactions.
Once molecules are formed, stable compounds can be incorporated into young disks as new YSOs condense out of the ISM.
This pathway is supported by measurements showing the organic composition of comets and the ISM to be quite similar
\citep{blw00,isc00}.

Laboratory results have shown that simple ices can be turned into a zoo of amino acids and other complex organics simply
by the presence of UV flux followed by a warming stage \citep{nab08,ncy07,bds02,mms02}.  Though these studies are
performed at fluxes higher than the ISM, their total photon dosages are comparable to the current problem.  Amino acids
similar to those seen in meteorites are created.  Moreover, chemical reactions (including photolysis) which occur at
cold temperatures $<70\ \rm{K}$ can preferentially increase D/H because deuterium's higher mass allows it to bond more
readily than H at low temperatures.  D/H enrichments are seen throughout the ISM, and the D/H enrichments in amino acids
suggests that they too have a low-temperature origin compared with the water in meteorite parent bodies \citep{sbd01}.
Continued UV exposure may sometimes destroy the same large molecules the UV created earlier, but for a proper range of
timescales and fluxes, cold-temperature photolysis of ices may be a more plausible pathway to forming the meteoritic
amino acids than warm aqueous synthesis.

The ISM formation model is not without problems.  First, the high optical depth within dense molecular clouds blocks
nearly all external UV light.  The only source of UV within the clouds is that caused by cosmic rays interacting with
gas in the clouds, resulting in a UV flux of $10^{-4}$--$10^{-5}$~G$_0$, where G$_0$ is the interstellar flux at the Sun
today \citep{pt83}.  Second, some organic molecules, such as the CHON particles in comets like Halley, may be easily
destroyed by shocks during infall, although some debate exists on this point \citep{vvd09,mws93,zg90}.  Finally, some of
the common meteoritic amino acids, such as $\alpha$-aminoisobutyric acid (AIB) and isovaline, have not yet been produced
by irradiation in the laboratory \citep{hmd08}, and it is not known whether this is due to these species only being formed
during Strecker synthesis \citep[as hypothesized by][]{egb01}, or simply the ice experiments not using the proper
initial mixtures.  Nevertheless, overall the ISM formation model paints a plausible picture and may well play a role in
the solar system's organic history.

This paper proposes a third pathway for the formation of these complex organic molecules, which has not been examined
previously.  In our model, the solar nebula forms within a large molecular cloud similar to those in Orion.  The Sun and
its disk form completely, and condense and begin to grow.  Within the next 1~Myr, nearby O and B stars turn on, bathing
the young disk in UV light.  These UV photons photo-evaporate gas from the disk \citep{tb05,jhb98}, but also irradiate
small ice grains exposed in the disk's outer skin layer.  The simple ices grains, such as H$_2$O, CO$_2$ and NH$_3$, are
exposed to the UV and begin to photolyze into more complex species.  Each individual grain is exposed only briefly to UV
light; they spend most of their time in the disk's dark, turbulent interior.  Grains continue to grow, gas which has not
formed planets is lost due to photo-evaporation and viscous loss, and within 5~Myr a gas-free debris disk is left with
its ices enriched in complex organics.  These organics can then be incorporated into comets, asteroids, and the planets.
Organics produced in this way could complement, and perhaps greatly exceed, those produced by the other two pathways.
This method is similar to the exogenic ISM production in that it relies on UV photochemistry, but at far high flux
($10^6$~G$_0$ vs.  $10^{-5}$~G$_0$), at warmer temperatures, for a much shorter time.  The model allows for subsequent
aqueous alteration as seen in the meteoritic record.

Work by \citet{rob02} proposed that the organic material was created by X-ray irradiation of the disk by the Sun
during its T Tauri phase.  They did not present a disk model, but used high-precision D/H measurements of organic and
non-organic material to show that D/H fractionation varied with heliocentric distance, as would irradiation.
\citet{rpr06} compared the meteoritic and interstellar D/H values, along with their C-H bond dissociation energies,
and concluded that the solar system's D/H enrichment was created \textit{in situ}, rather than inherited from the ISM.
They proposed that the young Sun might provide the necessary UV source; their model did not study the timescales or
fluxes involved.

The source of the Earth's organic inventory (as opposed to the Solar nebula's) is a parallel question which has
received some attention.  It is believed that although some organics were probably synthesized on the young Earth
\textit{in situ} by lightning or spark discharges \citep{mu59}, shock heating in the terrestrial atmosphere
\citep{cs92}, warm ocean vents \citep{cor90}, or any number of other terrestrial processes, far more were probably
delivered from external sources such as comets, asteroids, and interplanetary dust particles \citep{cs92}.  This paper
does not further address the origin of life or the Earth's inventory, except to acknowledge that increasing global
abundance of organics in the solar nebula probably results in increased delivery to the Earth as well.

This paper uses a simple model to describe the production of complex organics in UV-illuminated disks in a variety of
cases.  The results are necessarily general, and do not explain the abundances or species in one particular
sample or disk, but provides an initial assessment of the problem.  External UV photolysis has been ignored in almost
all previous models, yet it may be one of the most important sources of energy in both the young Solar System and other
proto-planetary disks.  The problem is set up in \S\ref{sect:background}, and the model described in
\S\ref{sect:model}.  \S\ref{sect:results} contains results, which are discussed in \S\ref{sect:discussion}.  Conclusions
are in \S\ref{sect:conclusions}, and Appendix~A contains a derivation of the simple grain growth model used here.

\section{Background}

\label{sect:background}

\subsection{Energy sources in the early solar nebula}

Various surveys have shown that the majority of stars are born in dense clusters of 300--$10^4$~stars, where massive O
and/or B stars can form \citep{apf06,ll03}.  Due to the presence of decay products from short-live radionuclides such as
$^{60}$Fe that are only produced in supernovae, our Sun is thought to have formed in such an environment
\citep{hdh04,th03}.  The effects of such an environment can be enormous, compared with the classical `closed box' view
of Solar System formation where any environmental effects are easily ignored.  Photo-evaporation by external stars can
remove the disks on Myr timescales or shorter \citep{tb05,mjh03,jhb98}.  Close approaches between stars may be
important, especially in the outer disk \citep{dbd08,apf06}.  The high gas densities in such clusters can cause late
infall of molecular cloud material onto disks after planetesimals have begun to form \citep{tb08}.  And, accretion of
ejecta from supernovae and massive stellar winds may contaminate the disk, changing its structure, composition, or both
\citep{tb10,csl98}.  

The effect of the local environment on solid-phase chemistry in the Sun's pre-planetary disk has received scant
attention.  Passing references can be found in \citet{bab07}, \citet{isc00}, \citet{thr00}, \citet{feg99},
\citet{pri93}, and \citet{vbd93}.  \citet{pf89} provides a complete review but is now somewhat eclipsed by more newer RT
models in the astrophysical literature (discussed below).  Their paper does not directly address the formation of
organics, but examines the energy budget for chemistry from various sources in the young solar nebula.  They find that
the largest energy source is the gravitational energy given off by the nebula as it collapses.  In addition to the
entire bulk disk heating, individual grains are heated and may sublimate ices as they fall into the nebula
\citep{vvd09,ler91}.  While this may cause some fractionation and possible hydrolysis of organic molecules, simply
warming ices from 10~K to 200~K for a few hundred seconds during the infall is not itself a pathway to form amino acids.
However, their second-largest source, shocks and lightning, certainly is.  They estimate the total usable energy for
reactions by these processes to be roughly $4 \times 10^{-6}$ that of the total gravitational collapse energy of the
disk.  Radionuclides (\eg, $^{26}$Al) were estimated to provide a few orders of magnitude again less energy.  They also
estimated energy from photochemistry, based on fluxes from both the Sun and interstellar sources.  They argued that
photochemistry in the solar nebula was unimportant, especially from from internal sources.  However, more sophisticated
recent models show that photochemistry can in fact be the dominant energy source.

\textit{Internal UV sources.}  Several recent models for disk formation consider UV radiative transfer from the central
star to the inner disk.  The  model of \citet{wkt09} handles photochemistry using a Monte Carlo method in a 2D disk of
mass 1 MMSN, surrounding a T Tauri star.  Over 70 chemical species are considered, including five ices, with a total of
nearly 1000 different chemical reactions.  A broadly similar model with fewer species is described by \citet{jdh07}.
The model by \citet{gh04} describes the UV photochemistry of 73 gas species in a 10~Myr old debris disk, putting it in a
much lower optical depth regime than the other models.  All three models are computed for a fixed age and do not evolve.
The grain size distribution in all is uniform across the disk.  All three models present sophisticated pictures of the
radiative transfer and photochemistry within the disks, allowing predictions to be made for abundances and line
strengths.  However, the model does not include ice photolysis, grain growth, differences in grain size across the disk,
or external illumination.  The UV field reaches $\sim 10^4 G_0$ inward of 10~AU, but this is still much lower flux than
the $10^6 G_0$ or more from external illumination.  

In calculating the energy budget of the nascent solar system, \citet{pf89} estimated that the the line-of-sight UV
optical depth through the disk to be ``7 million to 110 million \textit{orders of magnitude.}"  In their model, this
extreme optical depth would prevent solar-driven photochemistry virtually anywhere outside the central 0.35~AU.
However, they assumed a direct line-of-sight from the source was required, which \citet{gla93} points out is not the
case because solar Ly$\alpha$ photons can reflect off of interplanetary hydrogen far above the disk plane, thus
illuminating the disk by reflected solar light.  Considering the higher density and reflectivity of the interplanetary
medium (IPM) at the time, and the increase in early solar UV flux at $10^4\times$ or so over present values,
\citet{gla93} calculate that the young solar system was exposed to a solar UV flux roughly 10 times the present solar
value.  Scattering of solar Ly$\alpha$ photons off of winds, jets and infalling material can also be a large source of
UV onto the disk.  Similarly, \citet{hjl94} showed that young stars can create a thin atmosphere above their disk, and
this disk provides a source of indirect UV illumination which can drive photo-evaporation (and thus photolysis), at
rates 10--100 times lower than external illumination would, or $10^4$--$10^5$~G$_0$ \citep{mjh03}.  \citet{acp06,acp06b}
also looked at the case of photo-evaporation from the central star in \textit{flared} disks, which avoids the
line-of-sight problem in \citet{pf89}.  Combined with loss to inward viscous evolution, \citet{acp06b} calculated disk
loss timescales of several Myr, consistent with observation of disk dispersal in young T~Tauri stars.  None of the
papers by Gladstone, Hollenbach or Alexander considered the photochemistry effects of UV, but between these results and
the detailed RT models above, it becomes clear that photochemistry from the central star can be important.
Table~\ref{table:fluxes} lists approximate fluxes for the various sources.

\textit{External UV sources.} While the interstellar UV flux in Taurus-like regions is low ($\sim 1$~G$_0$ outside the
cloud, and $10^{-4}$~G$_0$ inside), stars in dense clusters such as the Orion Nebula Cluster (ONC) are subject to 
fluxes from O and B stars on the order $\sim 10^6$~G$_0$.  Ices in these clusters' disks are exposed to high
levels of UV irradiation, enhancing the disks' abundance of organic molecules.  In contrast to the internal sources, the
external sources are stronger by $10^2-10^4\times$, and illuminate the entire disk evenly rather than dropping off
with radial distance.

This paper examines only the effects of the external flux.  Existing models provide sophisticated treatments of the
internal UV flux and the gas photochemistry; our goal here is to understand the broad global effects of external flux
on solid-state chemistry, in preparation for a more detailed model.

\subsection{Energy budget}

A quick calculation shows the importance of external irradiation relative to other energy sources.  First, consider the
flux intercepted by a disk from its central star of luminosity $L$.  In a disk with a radius:half-height ratio of 10:1,
the fraction of flux intercepted by the disk is $\simeq 1/10$, and the total energy deposited at all wavelengths in time
$\Delta t$ is just

\begin{equation}
\esol \simeq {{1 \over 10} L}\ \Delta t.
\end{equation}

The total amount of energy available for chemistry through thermal heating, shocks, lightning, and so forth can be no greater
than the disk's total collapse energy from the cloud to disk inner radius $R_0$, given a disk mass $M_d$ and stellar
mass $M_s$, or 

\begin{equation}
E_{\rm{c}} = {{G M_{d} M_{s}} \over R_0}.
\end{equation}
Finally, the total UV energy absorbed by the disk from external stars is roughly

\begin{equation}
E_{\rm{UV}} = \sqrt(2) \pi R_d^2\ F_{\rm{ext}} \Delta t.
\end{equation}

Typical values for solar-mass stars in dense clusters are are $M_s = 1 \msol$, $M_d = 0.01 \msol$, $L = \lsol$, $R_d =
100\ \rm{AU}$, $R_0 = 5\ \rm{AU}$, $F_{\rm{ext}} = 10^6\ \rm{G}_0$, and $\Delta t = 1\ \rm{Myr}$.  Plugging in, we find the
ratio $E_{\rm{sol}} : E_{\rm{c}}: E_{\rm{UV}}$ is approximately $1:100:10^4$.  That is, \textit{in 1~Myr, the external
UV dose received by the disk exceeds by $100\times$ the entire direct energy input from the central star, and
the external UV dose exceeds by $10^4\times$ the entire collapse energy of the disk}. \footnote{If the opposite
were the case, photo-evaporation could not remove the disk.} Moreover, although thermal gradients can limit the
amount of energy available for chemical reactions \citep{pf89}, the UV energy is deposited directly into ice and dust
grains, where individual photons cause photolysis.  Therefore, the disk's dominant energy source is external flux, and
this energy source's effect on chemistry cannot be ignored.

\subsection{UV Photolysis of Ices}

The experiments of \citet[][see also \citealt{ure52}]{mil53} showed that electric discharges within an atmosphere can
rapidly create amino acids.  Since this time similar experiments have been repeated with different initial species,
temperatures, phases, and energy sources, finding the same general result that amino acids are relatively easy to
produce given sufficient energy.  In an astrophysical context, amino acids can be produced with energy sources from ion
irradiation, to physical shocks, to pyrolysis (heating), to UV irradiation \citep{bds02, cs92, mb88, bb75, mu59}.
According to \citet{hmd08}, ``one conclusion is that energetic processing of almost any organic ice that contains C, H,
N, and O probably results in the formation of amino acid precursors, which can be hydrolyzed to give the acids
themselves.''

Most interesting in the regions of cold space are laboratory experiments in the last decade involving UV irradiation of
ice mixtures.  For instance, experiments by both \citet{bds02} and \citet{mms02} started with thin ice mixtures
including H$_2$O, CH$_3$OH, and NH$_3$, chosen to simulate ISM conditions.  The micron-thick ice mixtures were formed at
15~K, and then irradiated for 30 minutes at $10^7$~G$_0$.  After warming, analysis in a gas chromatograph detected six
amino acids, including glycine and alanine.  Their laboratory irradiation corresponded to $10^7\ \rm{yr}$ in the
interior of a dense cloud ($10^{-4}$~G$_0$), or 12~hours in the Orion nebula ($10^6$~G$_0$).  The net photolysis
efficiency was around 0.25\% (complex organics produced per photon), and by the end of their runs the majority of their
C and N had been converted into new compounds \citep{bsa95}.  More recent experiments by \citet{nab08,ncy07} measured
upwards of a dozen different amino acids formed from irradiation of H$_2$O, CO$_2$, and NH$_3$. 

The authors of all three experiments hypothesized that long exposures of cold ices to UV light could explain the
multitude of organic molecules found in interstellar regions.  If the solar nebula collapsed from such an enriched
region of the ISM, subsequent incorporation of these species into preserved primitive bodies such as Murchison would be
a natural consequence of their birth environment.

Although UV can lead to the synthesis of organic molecules, excessive UV flux can be damaging to these same molecules,
as evidenced by terrestrial sales of skin-protection products.  \citet{bas04} looked into the UV destruction of amino
acids, and found that given enough flux they eventually formed nitriles, which are perhaps an order of magnitude more
stable against further destruction.  Similarly, \citet{ebd01} found that several hundred years of illumination at
1~G$_0$ destroyed most of the aminos in ice matrices.  \citet[][see also \citealt{gh06b}]{hmd08} showed that irradiation and
hydrolysis of nitrile-containing ices results in amino acids being created \textit{again}, so long-term irradiation may
eventually result in an equilibrium between aminos and nitriles. 

In Orion, the flux hitting a disk varies with time as a star orbits through the cluster, changing its distance and tilt
angle to the external O/B stars.  Typical distances from the Orion core range from about 1~pc to 0.01~pc, causing the
impinging flux to be in the range $10^3$ -- $10^7$ G$_0$ \citep{fa08,tb08,bsd98}.  $10^6$~G$_0$ represents the flux
hitting the disk skin.  However, young disks are very optically thick, so the vast majority of the ice grains are nearer
the midplane and thus sheltered from UV.  Simple calculations by \citet{thr00} based on the disk mass and small grains
gave typical initial optical thicknesses of close to $10^6$ at 10~AU; thus, these individual grains go through a
``broil-cool, broil-cool'' cycle as they circulate through the disk, only occasionally being exposed to the impinging
flux.  Disk midplane temperatures can be as warm as 50K out to 100~AU, so any nitriles produced during the irradiation
are likely to go on to form amino acids during when they shaded but still warm \citep{vvd09,wsw07,dch06}.

\section{Disk Model}

\label{sect:model}

The simple model used here is a modified version of that presented in \citet{tb05}.  The current model is 1D and
considers grain growth and external irradiation as a function of orbital distance $R$ from a 1~\msol\ star.  The disk is
azimuthally and vertically homogeneous, with initial parameters specified in Table~\ref{table:parameters}.  The initial
disk mass is 0.04~\msol.  The disk is made of gas and dust (ice), in an initial mass ratio of 100:1.  The disk has 40
logarithmically spaced radial bins, and it is evolved for $3\times10^5\ \rm{yr}$ using a self-adjusting time-stepper
coupled with a Crank-Nicolson integrator.  At the end of this time, photo-evaporation has largely removed the gas disk.
Real disks might continue to be exposed to UV for several Myr, but because of rapid grain growth, most of the
irradiation of small grains happens at the beginning of the simulation and the results are not strongly sensitive to the
time cutoff.

\subsection{UV Photolysis}

UV photons can create complex organic molecules or their precursors from simple ices.  The effect can be roughly
parametrized with a `photolysis efficiency' $\epsilon_p$, where $p \approx$~0.25\% in molecules per photon for
conversion from simple ices into amino acids.  This parametrization skips the complex chemical chains involved, but is
roughly the correct yield from pure ice \citep{bds02}.

At each timestep, the model computes the UV exposure onto the disk.  The instantaneous UV flux is

\begin{equation}
F(R,t) = {L \over {4 \pi d^2}} (1 - exp(-\tau(R, t)))
\label{eq:f}
\end{equation}
where the optical thickness $\tau(R,t)$ is

\begin{equation}
\tau(R,t) = {3 \over {4 \rho_d}} {\Sigma_{d}\, Q_{\rm{sca}} \over r(R,t)} ,
\label{eq:tau}
\end{equation}
assuming particles of size $r$ and density $\rho$.  The surface density $\Sigma_d$ is that of ice grains alone, and does
not change during the simulation.  We assume scattering efficiency $Q_{\rm{sca}} =1$, which is appropriate for all but
the very smallest grains and does not change the results.

At each timestep the flux $F(R,t)$ is computed, and the total number of photons intercepted by each bin is recorded.
These are then used to compute the total UV exposure in photons per molecule, 

\begin{equation}
\Phi(R,t) = \sum_t {{{F(R, t)} \over \Sigma_d} {1 \over m_m}} \Delta t, 
\label{eq:phi}
\end{equation}
where $m_m = 18\ \rm{amu}$ is the typical molecular mass.

The present model assumes a fixed external flux of $10^6$ G$_0$.  This flux is what the well-studied Orion proplyds at
0.1~pc receive, and is picked for easy comparison with previous results on those disks.  Two additional cases look at
higher and lower fluxes.

\subsection{Grain Growth}

Grain growth is important to the model, because once surface area is locked up in large grains, it cannot be exposed to
UV radiation.  Both observational and theoretical arguments shows that grains grow rapidly, reaching centimeter or meter
sizes on timescales as short as decades \citep{dbc07, jom07, tbe01}.  Grain growth (together with gas loss) causes the
disks to clear in optical depth from the inside out, typically within 1--5~Myr \citep[\eg][and references
therein]{aa07}.

The physical mechanisms for the early stages of grain growth are not well understood.  However, a simple
parametrization is useful for the present study.  \citet{thr00} performed semi-analytic calculations of accretionary
grain growth of a size distribution $n(r,R,t)$ of particles.  For grains up to a few cm, the grains were coupled to the
gas with collision velocities depending on the gas eddy speeds of \citet{mmv88} and \citet{vjm80}, while for larger
sizes they used particle-in-a-box collision probabilities.  In both regimes they assumed a constant sticking
probability $\epsilon_s$.  After integrating their distributions numerically, they combined an analytic expression for
grain growth with numerically-determined coefficients from the eddy velocity simulations to describe a the net grain
growth.  In their model, adopted here and derived in more details in Appendix~A, the typical grain size $r$ grows as

\begin{equation}
{dr\over{dt}}(R, t) = 1.2\times10^6\ e^{42.6\, p} R^{-2 - {{3\,p}\over2}}\ \epsilon_s^2\ \Sigma_{d1}^2\ \rho_d^{-5/3} \ t
\label{eq:rgrain}
\end{equation}
for time $t$ and orbital distance $R$. $\Sigma_{d1}$ is the initial dust surface density at 1~AU, and $\rho_d$ is the
dust grain solid density.  $\epsilon_s$ is the sticking efficiency, taken to be a constant in the range
0.001--1.  $p$ is the radial mass exponent, where $\Sigma \approx R^{-p}$.  The constant and exponential
are determined semi-analytically, and have units such that the final expression is correct for cgs inputs.  For
particles above one~meter, processes such as gravitational instability dominate over accretionary growth, and photolysis
is unimportant so growth is simply turned off.  Although eq.~~\ref{eq:rgrain} is a simple expression and considers only
a single characteristic size, it broadly agrees with more detailed calculations of grain growth in the micron-to-meter
range \citep[\eg,][]{wei97}.

\subsection{Photo-evaporation and Viscous Evolution}

The gas disk evolves using the standard prescription of \citet{pri81b}, with $\alpha = 0.01$.  For simplicity only gas
is transported, not dust.  Photo-evaporation by an external O star of brightness $10^5$~\lsol\ is also included; the
photo-evaporation model is as laid out in \citet{mjh03}.  The photo-evaporation model considers both EUV and FUV
flux from the external star as it removes the gas disk in $10^5$ -- $10^6$ years.  Flux from the disk's own central star
is not included.  The nominal stellar distance is 0.1~pc (giving a flux $10^6$ G$_0$); additional cases consider
distances of 0.01~pc and 0.5~pc.  Photo-evaporation removes gas but not dust, because except at the outer edge, grains
grow too quickly to be entrained in the escaping flow of gas \citep{thr00}.  In the current model, photo-evaporation
sets the timescale for removal of the gas disk, but does not itself affect photolysis.

Dense clusters such as Orion also are rich with outflows, stellar winds, and close stellar encounters.  These effects, while
easily visible in the nebula, have only minimal effect on individual disks \citep{tb08,amg07}, so they can be safely
ignored.

\section{Results}

\label{sect:results}

Simulations are presented here for our nominal case as well as six additional test cases, called \texttt{Massive},
\texttt{High Flux}, \texttt{Low Flux}, \texttt{Slow Grow}, \texttt{Delay}, and \texttt{Debris}.  The initial conditions
for each are listed in Tables~\ref{table:parameters} and \ref{table:varied}.

\begin{figure}
\centerline{\scalebox{0.6}{\includegraphics{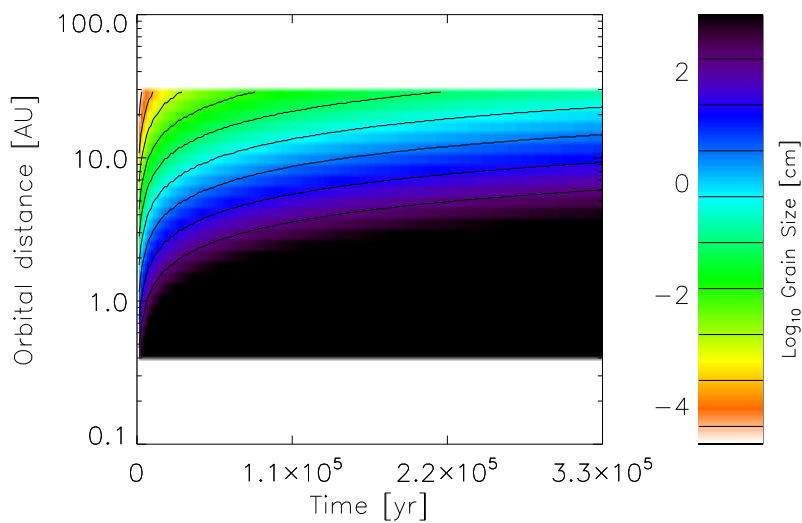}}}
\caption{Grain size, \texttt{Nominal}.  Grains grow to almost cm sizes throughout the disk within $3\times 10^5\ \rm{yr}$.
Growth is stopped at 5~m (black) as other growth mechanisms begin to take over.}
\label{fig:nominal_rgrain}
\end{figure}

\begin{figure}
\centerline{\scalebox{0.6}{\includegraphics{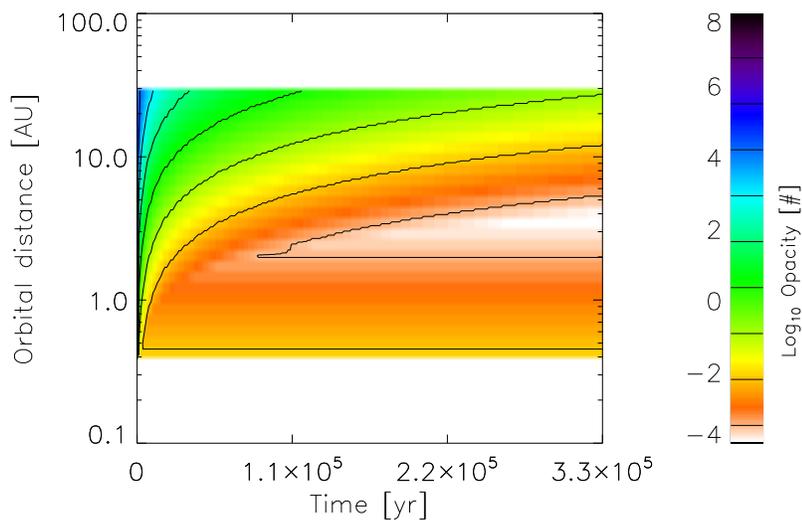}}}
\caption{Opacity, \texttt{Nominal.}  By the end, the opacity has dropped to below 1 throughout the disk as grains grow.
In the regions of low opacity, photolysis is inhibited because the disk intercepts little flux.  }
\label{fig:nominal_opacity}
\end{figure}

\begin{figure}
\centerline{\scalebox{0.6}{\includegraphics{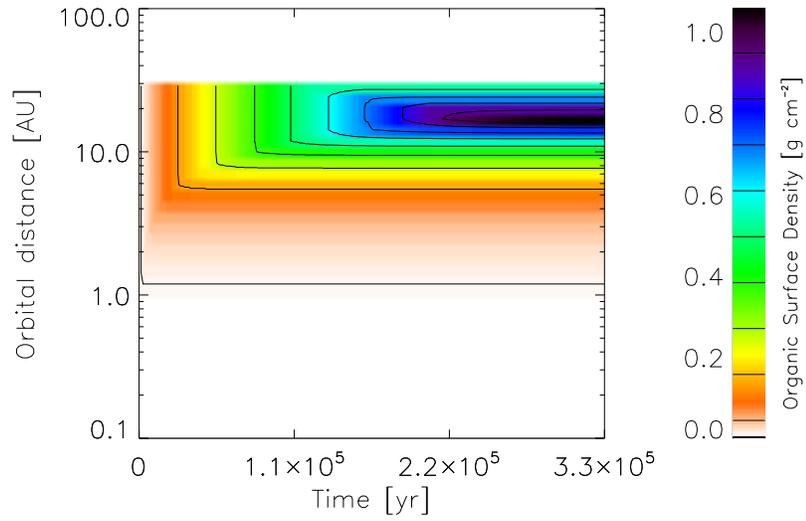}}}
\caption{Organic production, \texttt{Nominal}.  The integrated flux and opacity is converted into a maximum organic
photolysis yield.  The peak density is reached near the outer edge: inward of this grains grow rapidly, and outward the
disk density drops off.}
\label{fig:nominal_photolysis}
\end{figure}

\begin{figure}
\centerline{\scalebox{0.6}{\includegraphics{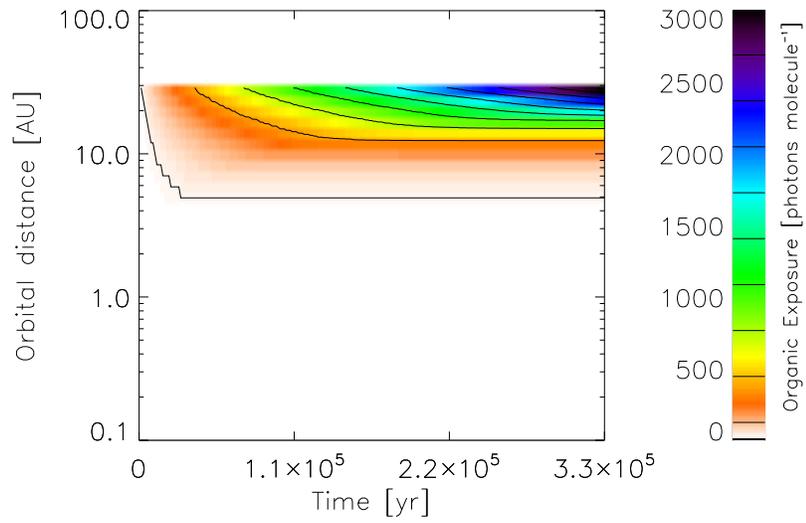}}}
\caption{Photons per molecule, \texttt{Nominal.}  This shows the net exposure of each original ice molecule in the disk
to UV flux.  Grains at the outer edge receive the highest flux because of their slow growth and the low surface density.}
\label{fig:nominal_photolysisppm}
\end{figure}

\begin{figure}
\centerline{\scalebox{0.6}{\includegraphics{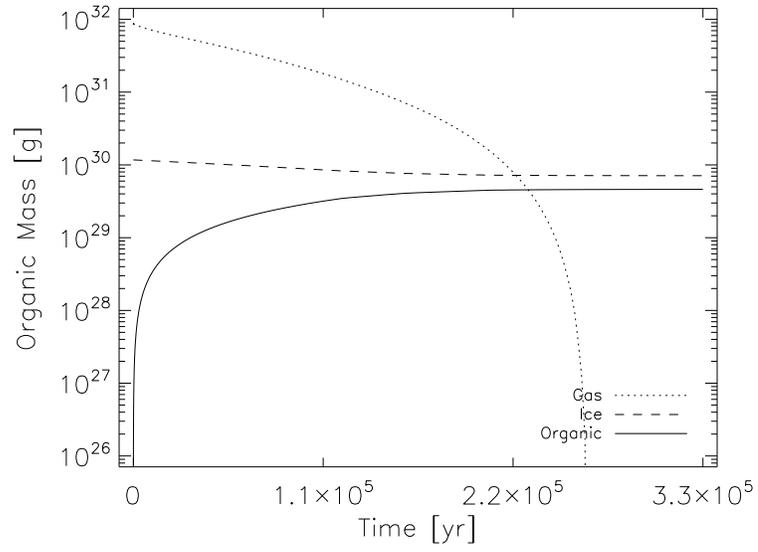}}}
\caption{Mass evolution, \texttt{Nominal.}  As ices are converted into organics, the solid and dashed lines approach
each other.  The dotted line plots the mass of the gas disk, as it is dispersed by photo-evaporation and viscous
evolution.}
\label{fig:nominal_photolysistot}
\end{figure}

Results for the \texttt{Nominal} case running for $3\times 10^5\ \rm{years}$ are shown in
Figs.~\ref{fig:nominal_rgrain}--\ref{fig:nominal_photolysistot}.  Grains grow steadily and reach meter sizes at the
inner edge and almost $1\ \rm{cm}$ at the outer edge.  As a result the opacity drops (Fig.~\ref{fig:nominal_opacity}),
reaching unity at the inner edge in about 10,000~years and dropping below this throughout the entire disk by $10^5\
\rm{yr}$.  The surface density at the inner edge is highest, but the rapid grain growth more than compensates for this,
clearing this region earliest.  At first, the production of organics proceeds at a uniform rate across the disk, because
all regions of the disk are optically thick and therefore intercepting every single photon
(Fig.~\ref{fig:nominal_photolysis}).  As the inner-disk opacity drops, photons begin to pass through the disk without
interacting.  This causes organic production to stop in this region, and this `production front' slowly moves outward in
the disk until organic production has ceased throughout, at about $2\times 10^5\ \rm{yr}$.  Further integration beyond
this point produces no more changes, because the disk's opacity has dropped below unity and little UV flux is
intercepted.  Looking at the total UV exposure (Fig.~\ref{fig:nominal_photolysisppm}), the greatest dose is delivered at
the outer edge, which receives $\sim3000$~photons~molecule$^{-1}$.  The inner region receives a much smaller dose of
$\lessim$100~photons~molecule$^{-1}$ inward of 5~AU.  The outer edge's dose is greater because of both slower grain
growth and lower surface density.  The dose of 3000 photons molecule$^{-1}$ is be sufficient to photolyze raw ices into
complex organics about seven times over.  As seen in Fig~\ref{fig:nominal_photolysisppm}, the peak organic density is at
$\sim$~20~AU.  Outward of this the disk's raw material inventory drops, and inward the grain growth is too fast to allow
for much photolysis.  By the end of the run, total organic production is $4.5\times10^{29}\ \rm{g}$, or close to half of
the original ice mass of $1.1\times 10^{30}\ \rm{g}$ (Fig.~\ref{fig:nominal_photolysistot}).  Photo-evaporation has
removed the entire gas disk in $2.5\times10^5\ \rm{yr}$. Numerical results from this run and all others are in
Table~\ref{table:varied}.

\begin{figure}
\centerline{\scalebox{0.6}{\includegraphics{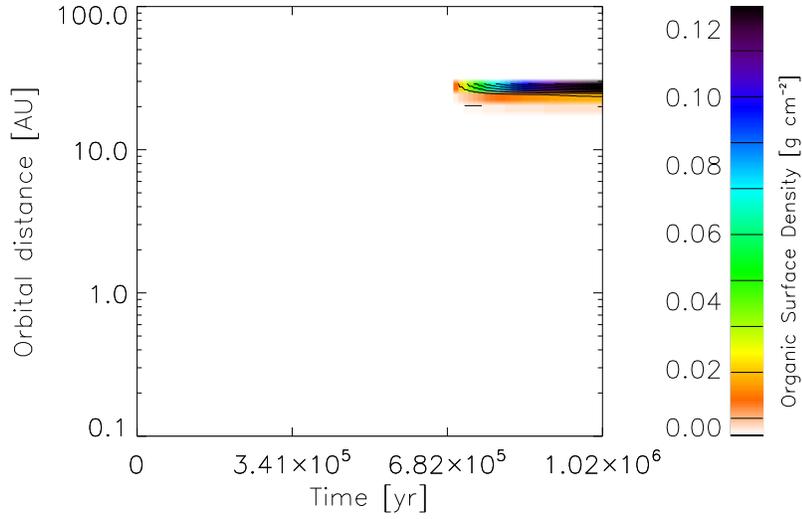}}}
\caption{Photolysis, \texttt{Delay.}  In this case the disk is given a head-start of 700,000 years to evolve before the
UV flux is started.  Grains grow, and the total organic production is about 5\% of that in the \texttt{Nominal} case.}
\label{fig:delay_photolysis}
\end{figure}

\begin{figure}
\centerline{\scalebox{0.6}{\includegraphics{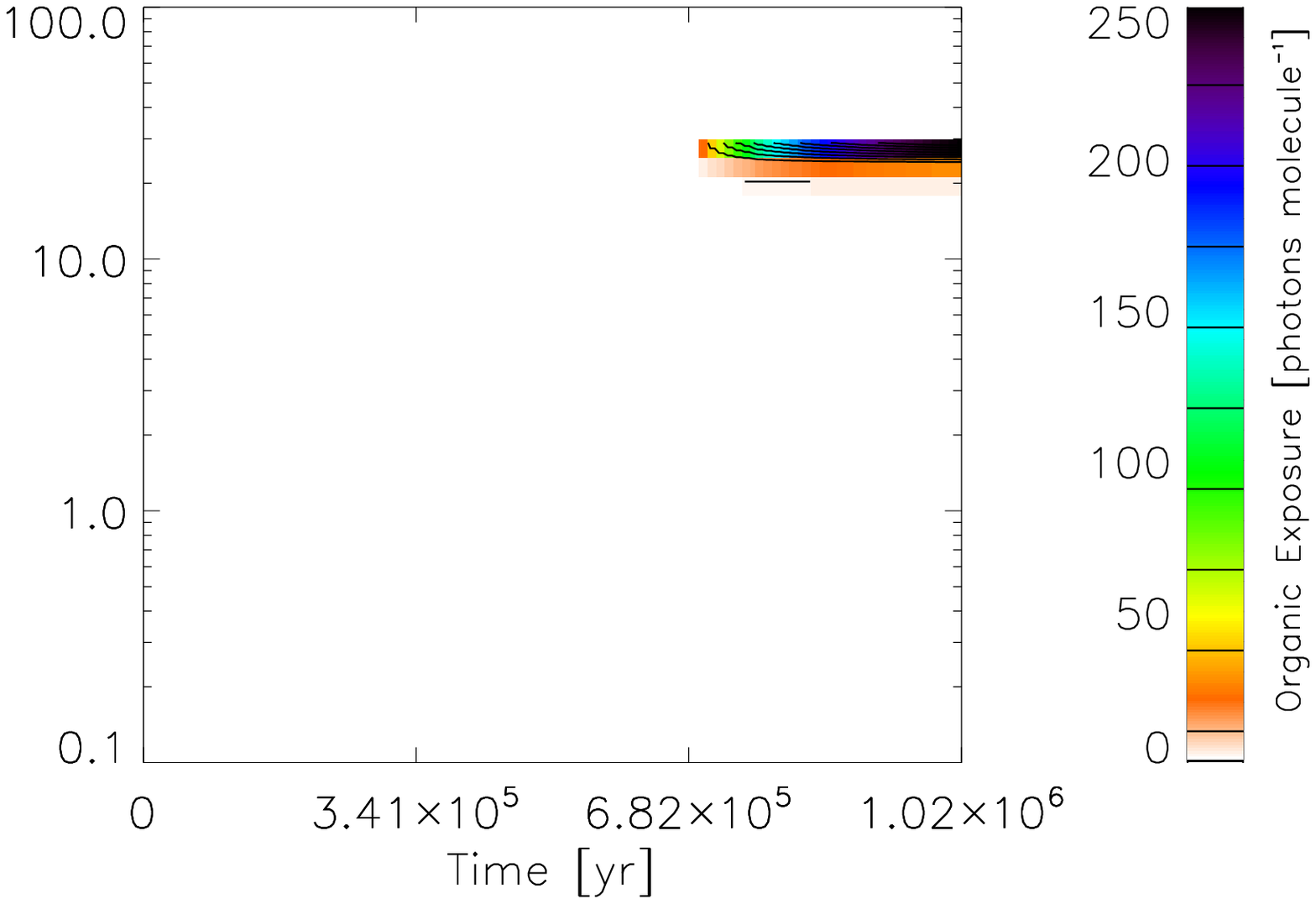}}}
\caption{Photons per molecule, \texttt{Delay.}}
\label{fig:delay_photolysisppm}
\end{figure}

\begin{figure}
\centerline{\scalebox{0.6}{\includegraphics{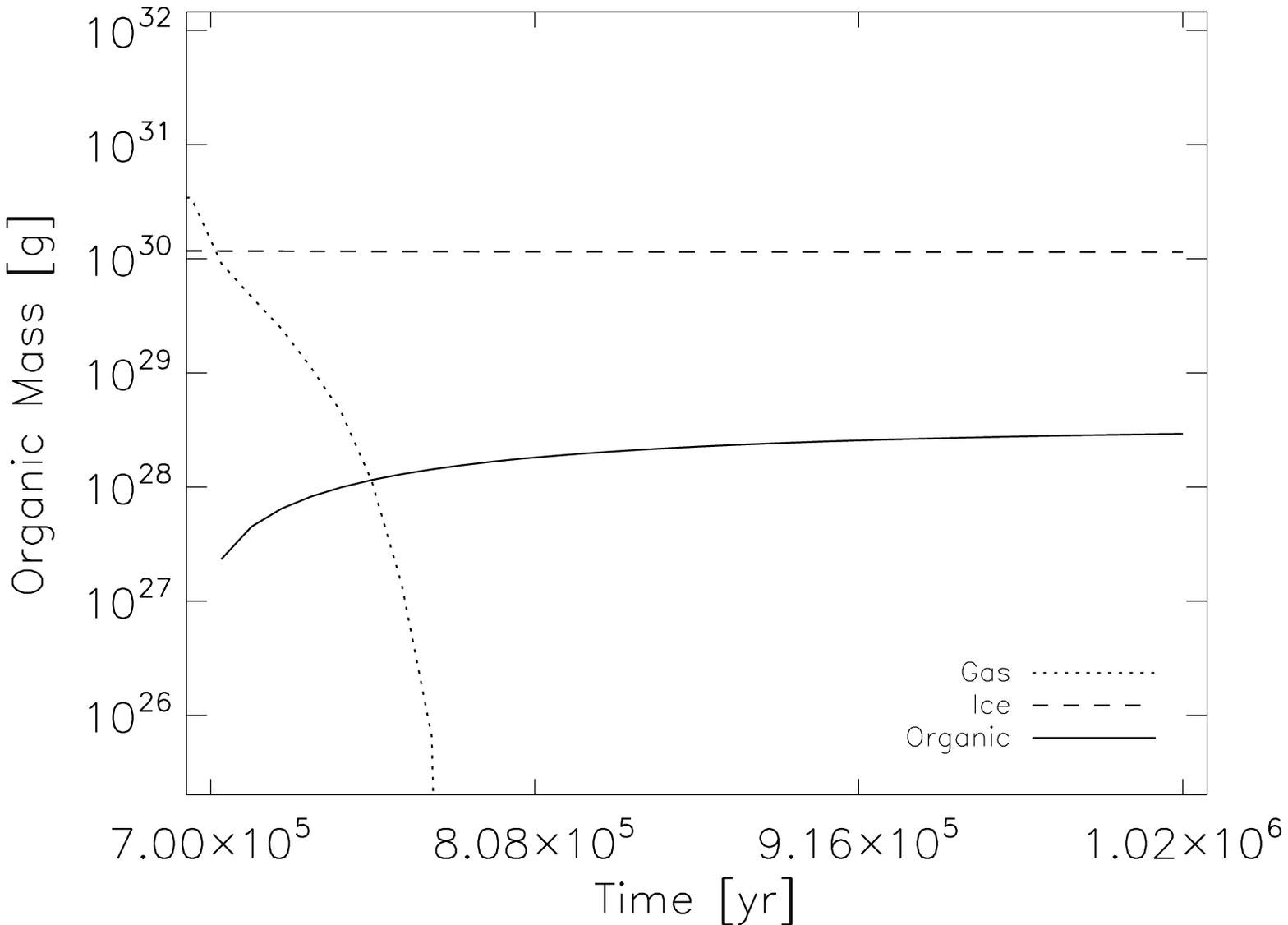}}}
\caption{Mass evolution, \texttt{Delay.}}
\label{fig:delay_photolysistot}
\end{figure}

Next shown (Figs.~\ref{fig:delay_photolysis}--\ref{fig:delay_photolysistot}) are results from the \texttt{Delay} case.  In
this, the disk is the same, but the UV source comes on only after a delay of $7\times 10^5\ \rm{yr}$, allowing for
some grain growth to happen before photolysis.  As can be seen, the resulting organic surface density and UV exposure
are about 10$\times$ lower than that of the \texttt{Nominal} case.

\begin{figure}
\centerline{\scalebox{0.6}{\includegraphics{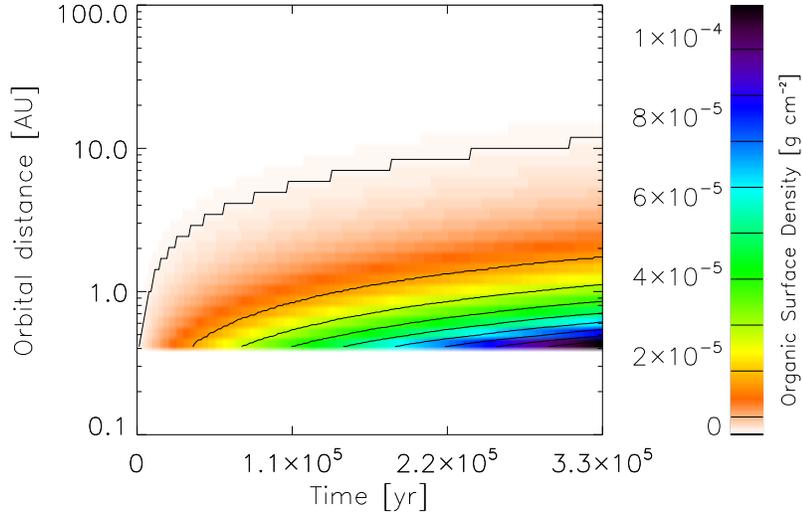}}}
\caption{Photolysis, \texttt{Debris.}  Here the grains start as a uniform 1~m size, with the same 
radial distribution as in the \texttt{Nominal} case.  Organic exposure is very low, about $10^{-6}$ that of
the \texttt{Nominal} model.}
\label{fig:debris_photolysis}
\end{figure}

\begin{figure}
\centerline{\scalebox{0.6}{\includegraphics{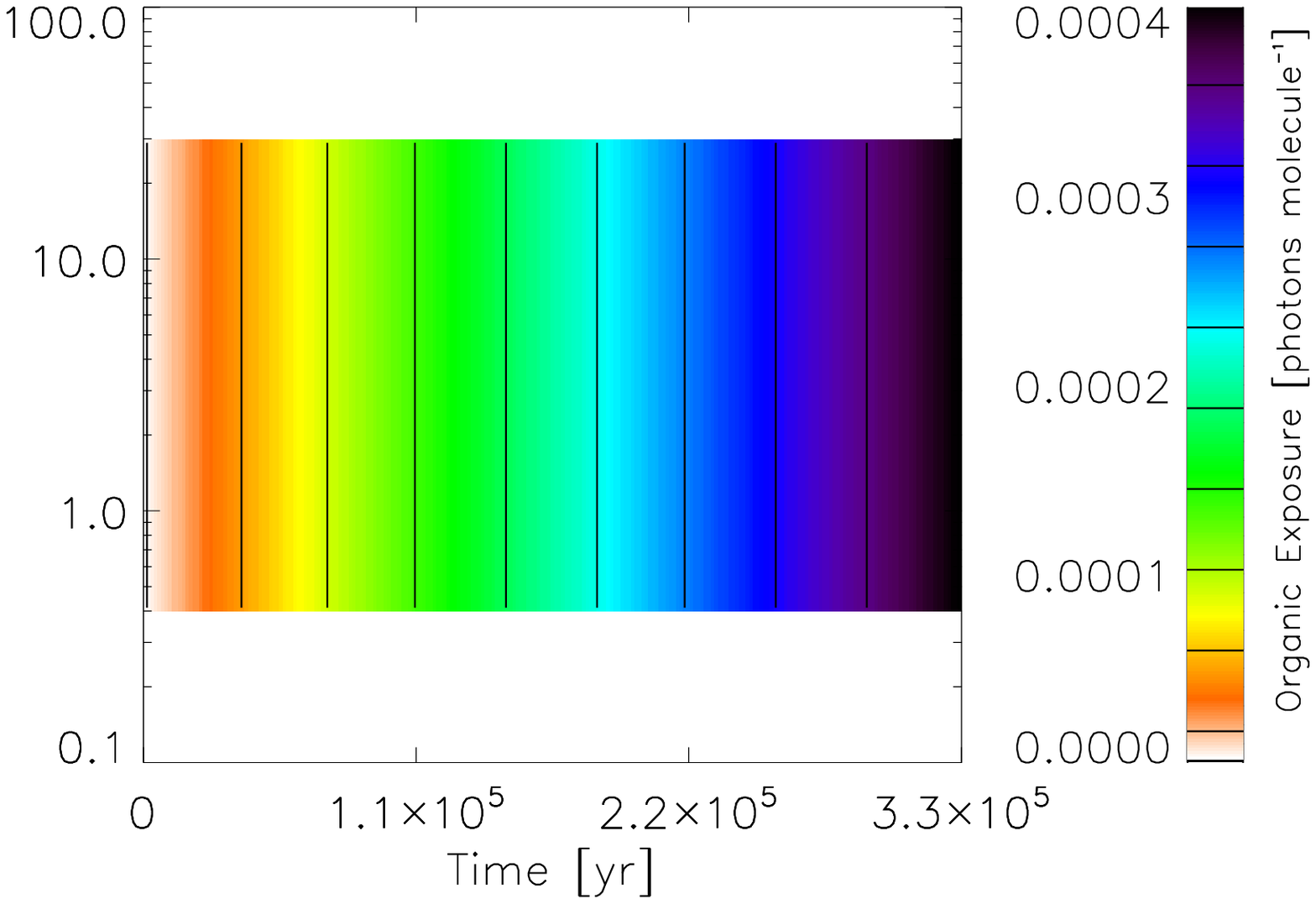}}}
\caption{Photons per molecule, \texttt{Debris.}}
\label{fig:debris_photolysisppm}
\end{figure}

\begin{figure}
\centerline{\scalebox{0.6}{\includegraphics{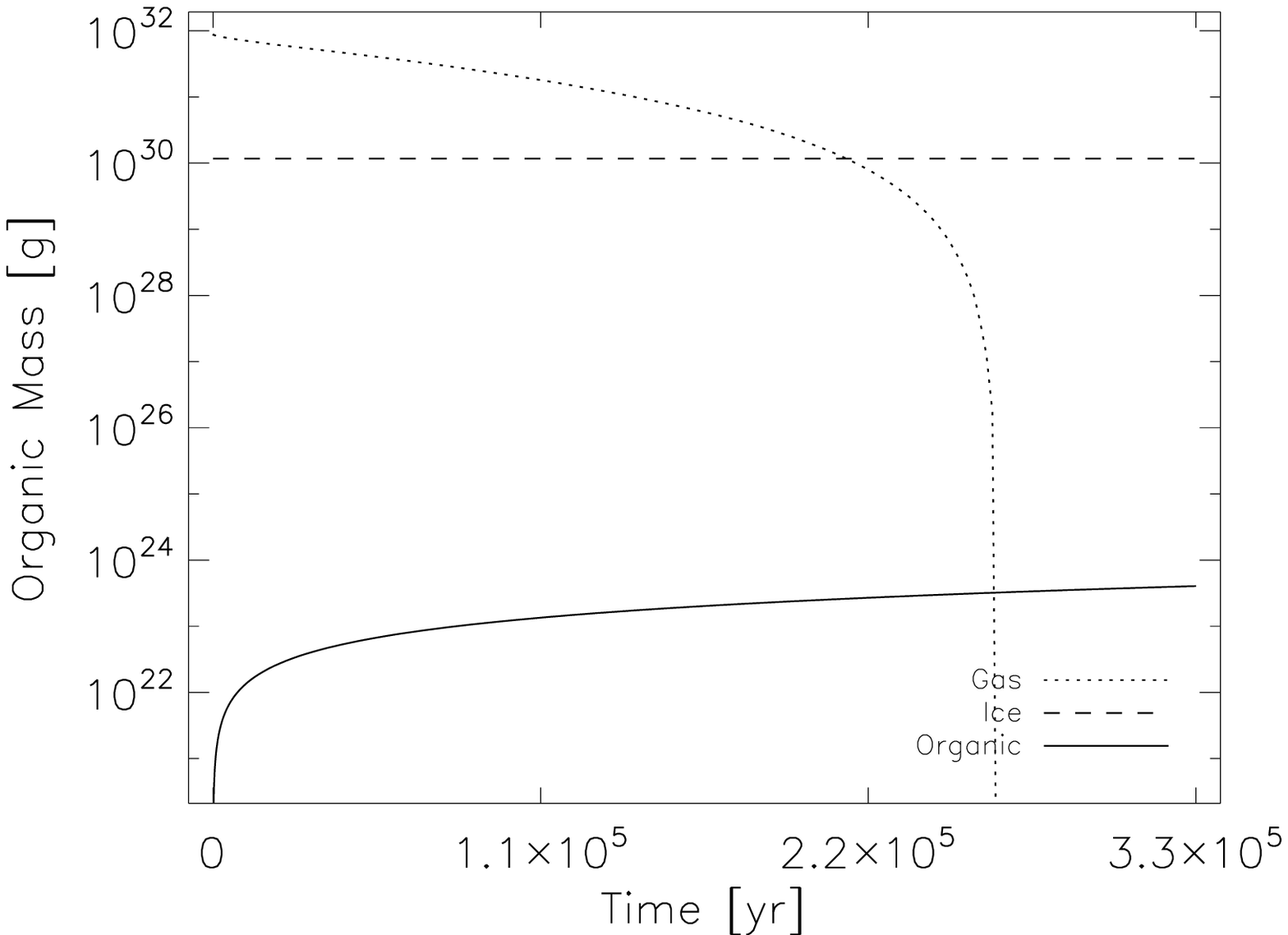}}}
\caption{Mass evolution, \texttt{Debris.}}
\label{fig:debris_photolysistot}
\end{figure}

Figures \ref{fig:debris_photolysis}--\ref{fig:debris_photolysistot} show the \texttt{Debris} case, where the grains have
been started at a constant size of $r = 1\ \rm{m}$.  In this case the disk's initial opacity is so low ($< 0.01$ at the
outer edge) that very little UV is intercepted, and the organic production is about $10^6\times$ lower than the
\texttt{Nominal} case.  Both the \texttt{Delay} and \texttt{Debris} cases simulate different conditions for older disks:
that is, disks which have evolved outside the \hii\ region and then enter into it, or ones born before the O/B stars
turn on.  Low-mass stars are generally believed to form over the course of several Myr, before the first high-mass stars
form, so most disks are probably somewhere between these two and the \texttt{Nominal} case.

\begin{figure}
\centerline{\scalebox{0.6}{\includegraphics{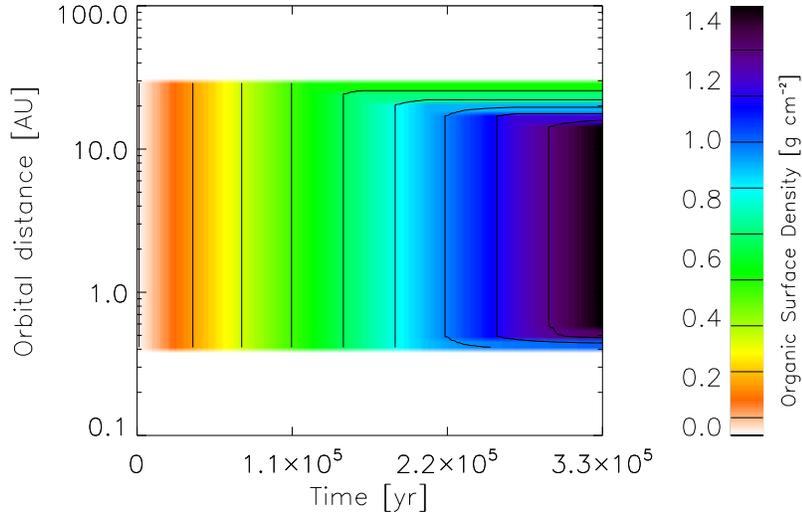}}}
\caption{Photolysis, \texttt{Slow Grow.}  The grain sticking coefficient is reduced from 1.0 to $10^{-3}$.  Grain growth
is slowed, and exposure to UV increases as a result because of long-lasting small grains.}
\label{fig:slowgrow_photolysis}
\end{figure}

\begin{figure}
\centerline{\scalebox{0.6}{\includegraphics{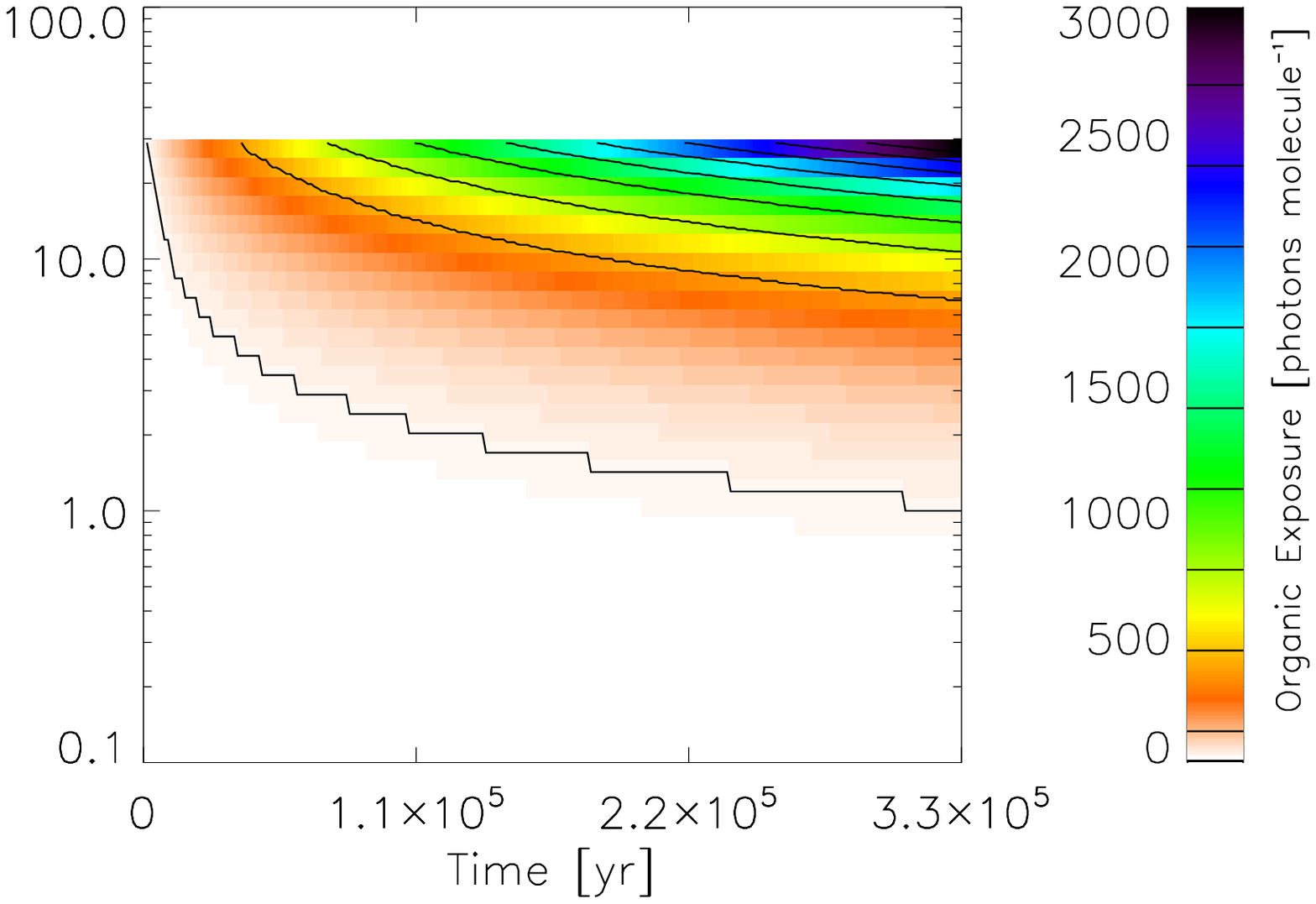}}}
\caption{Photons per molecule, \texttt{Slow Grow.}}
\label{fig:slowgrow_photolysisppm}
\end{figure}

\begin{figure}
\centerline{\scalebox{0.6}{\includegraphics{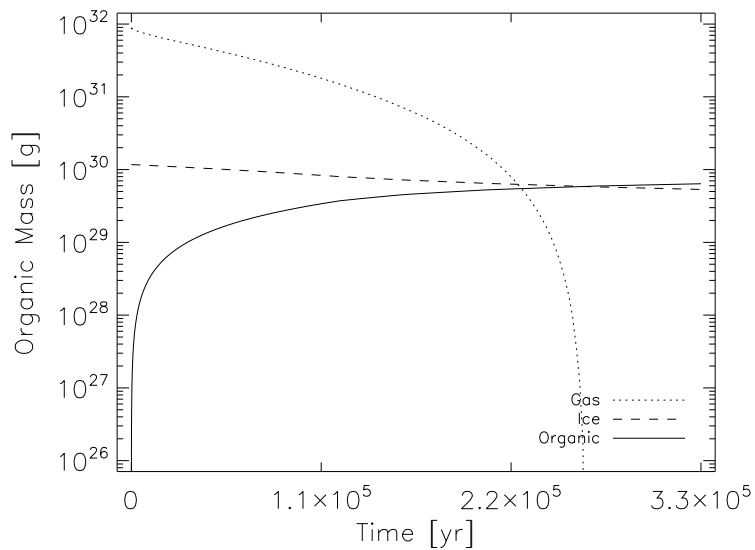}}}
\caption{Mass evolution, \texttt{Slow Grow.}  In this case more than 50\% of the original ices can be photolyzed into 
organic molecules.}
\label{fig:slowgrow_photolysistot}
\end{figure}

In the \texttt{Slow Grow} case (Figs.~\ref{fig:slowgrow_photolysis}--\ref{fig:slowgrow_photolysistot}), the grain
sticking efficiency has been reduced to $\epsilon_s=0.0001$, to promote slower grain growth and longer-lived disks.   As
a result the total UV dose is increased by an order of magnitude, allowing ices to be entirely photolyzed outward of
7~AU.  The total organic production is a few times greater than the \texttt{Nominal} case, and production would continue
to increase given more time (in contrast to the \texttt{Nominal} model).

Results for three additional cases are shown without figures, which are qualitatively similar to those presented.  In
the \texttt{Massive} run, the disk mass was increased by 10$\times$ to 0.4~\msol.  This decreases the net UV dose per
molecule in the inner disk, but in the outer disk where there is already more than sufficient flux in the
\texttt{Nominal} case, the result is a net greater production of organics.  In the \texttt{High Flux} case, the UV flux
is increased by 10$\times$ to $10^7$ G$_0$, which is the highest that most disks typically encounter for brief periods.
The effect is to increase photolysis, but by less than 10$\times$ because the outer disk is already flux-saturated.
Finally, in the \texttt{Low Flux} case, corresponding to a disk which is 0.5~pc distant from the cluster core, the
photolysis rate is scaled downward, roughly in proportion to the reduced flux.

Most of the runs share several similarities, including: a) the total raw exposure is sufficient to photolyze the entire
initial ice abundance of much of the disk; b) the photolysis is greatest at the outer edge; and c) the dose in a typical
cluster is high enough that in $10^5\ \rm{yr}$ the entire disk outward of 20~AU is photolyzed.

The model here is simplistic in several regards.  First, the laboratory yields used here are for photolysis
of virgin ice surfaces exposed for short times, and the efficiency $\epsilon_p$ will drop with time as
the virgin surface becomes irradiated.  More processing will also result in the destruction of some amino acids
into nitriles and gas-phase molecules \citep{bas04}, so the quantities presented here are upper limits.  Heating
from inward radial transport and shocks may cause additional destruction.  The total UV dose in the \texttt{Nominal} case
is high but not extreme: the highest-dose region at the outer edge receives only 7$\times$ the flux required for
complete photolysis, and regions inside of the edge receive far less than this ($<1$ inward of 10~AU).  While
eventual destruction of many of these organics must be assumed, the actual amount depends on parameters outside the
bounds of this study.

It is assumed that each individual grain is uniform throughout, while in reality flux is deposited only to the outer
skin of a grain, where it can build up a crust layer.  UV may be able to penetrate up to a few mm in rough or fractured
surfaces, but anything beyond this is doubtful (\citealt{ogj07} claims 1 meter penetration of UV into
Europa, but this seems highly unlikely in the case of any realistic ice).  However, most of the flux in the simulations
is deposited when the grains are less than a few mm, making this a reasonable approximation.  For instance,
Fig.~\ref{fig:nominal_photolysis} shows that UV deposition has stopped by about 20,000 years in the \texttt{Nominal}
disk.  Fig.~\ref{fig:nominal_rgrain} shows that by this time, grains in this region remain mm-sized.  In addition,
grains of this size are likely to be fluffy, cracked, collisional aggregates which expose new surface regularly.  And in
all cases, the rapid grain growth occurs in the inner disk, where the models predict very little photolysis in the first
place.  The handling of UV deposition here is thus probably accurate enough given the model's other uncertainties.

The model here assumes the disk to be made of ice only, and not silicates.  In reality the ice:silicate ratio is
approximately 1:1 outward of the snow line, causing the net flux to be roughly halved for homogeneously mixed grains.
Even within individual ice grains, the formation process may concentrate some reactants toward the center and others
toward the outside, slowing photolysis further \citep{cdf03}.

In addition to chemically processing grain surfaces, UV photons can erode grains by photo-sputtering molecules into the
gas phase, where they are easily removed by the photo-evaporative wind.  Sputtering may be as high as 1~meter per Myr
\citep{thr00,wbj95} for fresh surfaces at $10^6~$G$_0$.  However, for both photolysis and sputtering, large bodies will
probably build up a crust of silicates which will prevent further effects of exposure.  It is difficult to estimate this
effect without more detailed laboratory modeling of heterogeneous mixtures of thick, collisionally processed ices and
silicates.  The interaction between sputtering and photolysis should be considered in future models.  

Our model handles the radial transport of gas but not dust.  Dust transport is a poorly understood yet important
process \citep[\eg][]{bta06}, and should be included in the next generation of models.

The model here finds that the disks become optically thin by $10^5$--$10^6\ \rm{yr}$, somewhat shorter than the observed
disk lifetimes of 1--5~Myr.  The difference between these two timescales is that our model considers just monotonic
grain growth, while real disks continue to release small secondary dust grains during particle collisions.  These
continually produced small grains remain visible even as most of the mass of the disk grows into planetesimals.  UV
photolysis (and photo-evaporation) can therefore continue for much longer periods than the lower-limit timescales
predicted here.  The assumption of short disk lifetimes here causes our flux numbers to be conservative.

\section{Discussion}

\label{sect:discussion}

The model here leads to one robust result: there was sufficient external UV flux in the young solar system to allow for
ice photochemistry to play an important role in the solar system's chemical makeup.  Complex organics can be rapidly
produced in flux environments typical of dense star clusters, especially in the outer solar system.  Previous models for
photochemistry in our solar system have not considered ice photochemistry from external UV sources.  The current model
describes a new pathway for production of complex organics in the young solar system.  This adds to the two existing
pathways of aqueous alteration on meteoritic parent bodies and photo-synthesis in the ISM, thus relaxing the conditions
required in the early solar nebula.  The total external UV energy available can exceed that of all other
energy sources combined.

Of the simplifications made here, two are worth discussing.  First, photolysis is assumed to be a one-way process,
when in reality continued UV can break apart amino acids that it creates.  Continued laboratory work and improved
modeling will constrain the net production rate, allowing improvements from the upper limit yields presented here.
Second, the interaction between photo-evaporation and photolysis is simplified.  Photo-evaporation is driven by UV
absorption into dust or ice grains, and if they are tiny enough the photo-evaporative outflow can entrain these grains,
as seen in some large outflows \citep{brm08}.  Small grains might be continually produced as collisional debris in young
disks.  So, if the disks are dominated by a continually refreshed population of small dust grains, these grains might act
as a shield, protecting the inner disk from UV but themselves providing a large source of irradiated organics delivered
back to the ISM in the outflow.

The largest free parameter in these models relates to timing: if disks evolve in a dark environment for more than
$10^5$--$10^6$~yr, grains grow to meter sizes and larger where UV has no chance of penetrating to cause any more than
trivial amounts of photolysis.  In dense star clusters, the formation of low-mass stars is thought to continue steadily
up until the moment that massive O/B stars form.  Once they have turned on and the molecular cloud is ionized, star
formation stops.  Low-mass stars that form just prior to the O/B stars will thus end up with the largest amount of
organics, while the oldest disks (up to several Myr in the case of Orion) will have much less.

The results here predict that organic molecules will be abundant in young circumstellar disks, but to date only a few of
these molecules have been actually detected.  However, help is on the way in the form of the Stratospheric
Observatory For Infrared Astronomy (SOFIA) and the Atacama Large Millimeter Array (ALMA).  SOFIA, operating in the
3--300\ \micron\ range at a spectra resolution of $\sim$ 100,000, will be able to detect vibrational and bending lines
of water ices, methanol, and some CO bands.  Organics such as amino acids are too heavy to have easily visible bands,
but the structure of the water ice matrix as modified by these heavier molecules will be able to be studied by SOFIA.
ALMA will provide brand new 0.1" high resolution imaging of the disks, allowing insight into the radial distribution of
molecules across the disk.

\section{Conclusions}
\label{sect:conclusions}

\b There was sufficient flux from nearby O/B stars in the Sun's birth environment to convert the the majority of the
ice mass in the young Solar System and other proto-planetary disks into complex organic molecules within a few $10^5\
\rm{yr}$.  Even when considering the disks' high optical depth and rapid grain growth, both of which sequester surface
area, enough ice remained exposed for photolysis to continue rapidly.  

\b UV photolysis from external sources is likely to be the source of the meteoritic amino acids, which have D/H ratios
difficult to reconcile with the warm, aqueous environments required from Strecker synthesis of these species.

\b Organic production is predicted to be substantially less for disks that have evolved for more than
$10^5$--$10^6$~years before O/B stars have turned on, because the grain growth in this time locks up surface area
against photolysis.

\b Externally driven photochemistry is most important in the outer disk, where grain growth is slow.  The inner disk is
warmer, more massive, and sees more rapid growth, and thus processes such as Fischer-Tropsch reactions may dominate the
production of organics here.

\b External UV flux in 1~Myr provides roughly $100\times$ as much flux as from the central star, and $10^4\times$ as
much energy as gravitational collapse.  This energy is easily usable for chemical reactions because it is directly
incident on ice grain surfaces.

\b Photo-evaporation can occur at the same time as photolysis of ices, but these two processes operate largely
independently.  Photo-evaporation removes only the smallest grains, and photolysis can continue long after the gas disk
has been removed.

Dense star clusters have historically been thought of as a hazard to planet formation, because the rapid timescales of
disk destruction limit the conditions under which planets can form.  Our earlier results showed that under some
conditions photo-evaporation may in fact \textit{speed} the formation of planetesimals \citep{tb05}.  Our new results
here add to this irony, showing that planetary systems that can form in such `hostile' environments may also be be among
the richest in organic molecules and and the pre-cursors of life.

\section{Acknowledgments}

This work was supported by NASA Exobiology grant NNG05GN70G.  I thank Max Bernstein, John Bally, and Randy Gladstone
for useful discussions.


\appendix

\section{Appendix: Grain Growth}

\citet{thr00} and \citet{tbe01} developed a simple model for accretionary grain growth.  The details of that model were
presented only schematically, so a more detailed derivation is given here.  In this model, particle grow by pure
accretion, which is often considered to be the dominant process for the earliest phase of growth, where particles range
from microns to $\sim 1$ meter.  

Consider a disk which has collapsed from the ISM, with an initially uniform particle size of $r_0$.  The disk is
arranged such that its full height $H$ at radius $R$ is given by 

\begin{equation}
H = R/10
\label{eq:appendix_h}
\end{equation}
Within this disk, dust grains collide and begin to grow by simple accretion.  For simplicity, we assume that the
particles can be defined by a single grain size $r(R)$ which is a function of orbital distance.  A
particle-in-a-box collision model gives the grain growth rate as

\begin{equation}
dm/dt = \sigma\, n\, v\, m\, \epsilon_s
\end{equation}
where the collision cross-section is $\sigma = 2\pi r^2$, $n$ is the particle number density, $v$ is the collision
velocity, the particle mass is $m = {4\over3} \pi r^3 \rho_d$,  and $\epsilon_s = [0,1]$ is the sticking efficiency for
a single collision.  Assuming vertical homogeneity and no radial transport, the particle volume density $n$ can be
computed directly from the disk structure as

\begin{equation}
n = \Sigma / (H\,  m)
\end{equation}
and the surface density of the dust component $\Sigma_d$ (normalized to $\Sigma_{d1}$ at 1~AU),
\begin{equation}
\Sigma_d = \Sigma_{d1}\ \left[ {R \over {1\, AU}} \right]^{-p} .
\end{equation}

To compute the collision velocity $v$, \citet{thr00} examined the interparticle velocities present in the convecting,
turbulent disk.  The collision velocity depends on particle size because particles are trapped in eddies of different
strengths.  Performing semi-analytic fits to the work of \citet{vjm80} and \citet{mmv88}, they approximated the
collision velocity for dust grains in a gas disk surrounding a 1~\msol\ star as

\begin{equation}
v = 10^2\, r^{1/2}\, ({2.2\, 10^{-16}\, R})^{p/4}
\label{eq:appendix_v}
\end{equation}
where all units are cgs.  Combining these eqs.~\ref{eq:appendix_h}--\ref{eq:appendix_v}, integrating, and solving for
$r$ yields the result

\begin{equation}
dr/dt = 1.2\ \times 10^6\, e^{42.6\, p}\, R^{- 2 - {{3p}\over2}}\, \epsilon_s^2\ \Sigma_{d}^2\ \rho_d^{-5/3}\, t
\label{eq:appendix_drdt}
\end{equation}
which can then be easily used to compute $r(t)$ or $m(t)$.  The constant and exponential out front are determined
semi-empirically and have the proper units such that the result is correct for cgs inputs.  The expression shows a
strong exponential dependence on the surface density exponent $p$, but this is deceptive: $p$ usually has only a small
range of values (1--2), and the dependence is reduced by other terms so that $p$ only has an appropriately small effect
on the grain growth.

\begin{figure}
\centerline{\scalebox{0.6}{\includegraphics{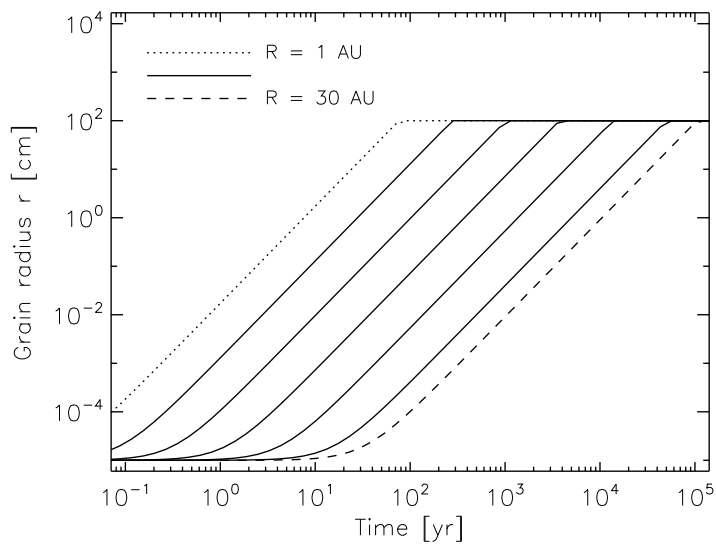}}}
\caption{Grain size evolution.  Plot show accretionary grain growth according to eq.~\ref{eq:appendix_drdt}, for disk of
$p = 3/2$ and 0.1~\msol; growth is stopped above 1 meter radius.  Radial bins are spaced logarithmically.}
\label{fig:drdt}
\end{figure}

\citet{thr00} tested this equation (Fig.~\ref{fig:drdt}) and found it to agree well with accretionary growth
calculations done numerically with full size distributions, using both their own calculations and those of \citet{mmv88}
and \citet{wei97}.  In the case of $\epsilon_s = 1$, it gives an upper limit to the grain size in a young disk growing
by coagulation.  For a standard 0.05~\msol\ disk with gas:dust ratio of 100:1 and 0.1~\micron\ initial grains,
eq.~\ref{eq:appendix_drdt} predicts growth to 1~cm in $10^4$ years and 1~m in $10^5$ years inward of 30~AU.  This rapid
growth is verified by \citet{wei97}, which finds growth of up to several meters within $10^5$ years at 30~AU. Because
the model considers only accretionary growth, it is useful only for particles up to a few meters in size, after which
gravity and settling begin to become important.  It also ignores radial drift, fragmentation, and possibly faster
mechanisms such as gravitational instability \citep{jom07}.  Nevertheless, for computing rough particle size estimates
and their dependence on disk parameters in the youngest disks, it provides useful limits which are consistent with more
complex codes.

%


\bibliography{papers}

\vfill \eject



\begin{table}[h]
\begin{tabular}{|l|l|} \hline
\textbf{Parameter} & \textbf{Value, \texttt{Nominal}}  \\
\hline \hline
Surface Density       & $\Sigma \propto R^{-3/2}$ \\ \hline 
Initial grain size    & $r_0$ = 0.2 \micron \\ \hline
Photolysis efficiency & $\epsilon_p$ = 0.0025 \\ \hline
Sticking efficiency   & $\epsilon_s$ = 1.0 \\ \hline
Stellar mass          & $M = 1 \msol$ \\ \hline
Initial disk mass     & $M_{\rm{0d}}$ = 0.04 $\msol$ \\ \hline
Disk edges            & $R_0$ = 0.4 AU; $R_1$ = 30 AU \\ \hline
Ice:Gas mass ratio    & 1:100 \\ \hline
UV flux               & $F = 10^6$ G$_0$ \\ \hline
UV start time         & $t_{\rm{UV}}$ = 0 yr \\ \hline
Run time              & $t_{\rm{run}} = 3\ 10^5$ yr \\ \hline
\end{tabular}
\caption{List of parameters, \texttt{Nominal} case.}
\label{table:parameters}
\end{table}


\begin{table}[h]
\begin{tabular}{|l|l|l||} \hline
\textbf{Trial Name} & \textbf{Changed Parameter} & \textbf{Organic Production} \\
\hline \hline
\texttt{Nominal}     & --                                       & $4.5\ 10^{29}\ \rm{g}$ \\ \hline
\texttt{Massive}     & $M_{\rm{0d}} \rightarrow 0.4~\msol$      & $1.6\ 10^{29}$ g       \\ \hline
\texttt{High Flux}   & F $\rightarrow 10^7$ G$_0$               & $7.0\ 10^{29}$ g       \\ \hline
\texttt{Low Flux}    & F $\rightarrow 2.5\times10^4$ G$_0$      & $3.1\ 10^{28}$ g       \\ \hline
\texttt{Slow Grow}   & $\epsilon_s \rightarrow 0.001$           & $6.3\ 10^{29}$ g       \\ \hline
\texttt{Delay}       & t$_{\rm{UV}} \rightarrow$ 700,000 yr     & $2.9\ 10^{28}\ \rm{g}$ \\ \hline
\texttt{Debris}      & r$_0 \rightarrow 1$ m                    & $4.0\ 10^{23}\ \rm{g}$ \\ \hline
\end{tabular}
\caption{List of parameters varied.}
\label{table:varied}
\end{table}

\begin{table}[h]
\begin{tabular}{|l|l|l|l|} \hline
\textbf{Source} & \textbf{Flux at 10 AU} & \textbf{Reference} \\ \hline \hline
UV flux inside dense molecular cloud     & $10^{-4}$--$10^{-5}$ G$_0$  & \citet{pt83} \\ \hline
Present day, interstellar                & 1 G$_0$                     & \citet{hab68}\\ \hline
Present day, Solar                       & 10 G$_0$                    & \citet{cox00} \\ \hline
Young Sun, reflected from IPM            & 1000 G$_0$                  & \citet{gla93} \\ \hline
Central T Tauri onto flared disk         & 3000 G$_0$                  & \citet{aa07} \\ \hline
External stars, small cluster            & $10^3$--$10^5$ G$_0$        & \citet{apf06} \\ \hline
External stars, large cluster            & $10^5$--$10^7$ G$_0$        & \citet{jhb98} \\ \hline
\end{tabular}
\caption{UV fluxes in different environments.}
\label{table:fluxes}
\end{table}

\clearpage



\clearpage


\clearpage


\clearpage


\clearpage


\end{document}